\documentclass[twoside]{article}

\usepackage[accepted]{aistats2020}
\usepackage{amsmath, amstext}
\usepackage{graphicx,psfrag,epsf}
\usepackage{enumerate}
\usepackage{microtype}
\usepackage{bbm}
\usepackage{bm}      
\usepackage[ruled,vlined]{algorithm2e}
\usepackage{algorithmic}
\usepackage{booktabs}
\usepackage{graphicx}
\usepackage{latexsym, amsbsy, amssymb}
\usepackage{hyperref}
 \usepackage[ruled,vlined]{algorithm2e}
 \usepackage{xcolor}
 \usepackage{mathtools}

\usepackage[round]{natbib}

\newtheorem{theorem}{Theorem}
\newtheorem{assumption}{Assumption}

\def\vt{\mathbf \Theta}
\def\vs{\mathbf \Sigma}
\def\vecX{\textit{\textbf{X}}}
\def\vect{\textit{\textbf{t}}}
\def\Plus{\texttt{+}}
\def\Minus{\texttt{-}}
\DeclareMathOperator*{\argminB}{argmin} 
\DeclareMathOperator{\poly}{poly}

\begin{document}

%

%

\twocolumn[

\aistatstitle{Inference of Dynamic Graph Changes for Functional Connectome}

\aistatsauthor{ Dingjue Ji$^\dagger$ \And Junwei Lu$^\sharp$\And  Yiliang Zhang$^\ddagger$\And Siyuan Gao$^\S$\And Hongyu Zhao$^{\dagger,\ddagger}$ }

\aistatsaddress{ Interdepartmental Program of Computational Biology \& Bioinformatics$^\dagger$, \\ Department of Biostatistics$^\ddagger$, Department of Biomedical Engineering$^\S$, Yale University  \\Department of Biostatistics, Harvard T.H. Chan School of Public Health$^\sharp$} ]

\begin{abstract}
Dynamic functional connectivity is an effective measure for the brain's responses to continuous stimuli. We propose an inferential method to detect the dynamic changes of brain networks based on time-varying graphical models. Whereas most existing methods focus on testing the existence of change points, the dynamics in the brain network offer more signals in many neuroscience studies.
We propose a novel method to conduct hypothesis testing on changes in dynamic brain networks. 
We introduce a bootstrap statistic to approximate the supreme of the high-dimensional empirical processes over dynamically changing edges. Our simulations show that this framework can capture the change points with changed connectivity. 
Finally, we apply our method to a brain imaging dataset under a natural audio-video stimulus and illustrate that we are able to detect temporal changes in brain networks. The functions of the identified regions are consistent with specific emotional annotations, which are closely associated with changes inferred by our method. 
\end{abstract}
\section{INTRODUCTION}
The time series of multivariate variables with underlying time-varying graphical structures are commonly studied in various areas such as transcriptomics \citep{rodius2016analysis, willsey2013coexpression}, finance \citep{silva2015modular, isogai2017dynamic}, and neuroscience \citep{kabbara2017dynamic, soreq2019dynamic}. With time-evolving gene co-expression networks, dynamic stock price networks, and dynamic brain connectivities in cognitive processes, these studies have shown the necessity of investigating dynamic changes in graph structures, especially when temporal changes or specific turning events are expected. 
\subsection{Motivation}
The turning events during a time course can be abrupt, and bring sudden perturbations to the system. For example, bifurcation exists in important biological regulatory networks such as the p53 system \citep{hat2016feedbacks}, where a small change or perturbation can cause a qualitative change of the state in the system \citep{champneys2013dynamical}. Structural breaks may be observed in other data types with a similar design or sudden changes caused by unexpected events, such as financial data during bursts of emergencies \citep{pepelyshev2015real}. 

Brain imaging is also a typical example of signals with graphical variety during a certain external stimulus, and the brain transits to distinct network states according to task demands \citep{gonzalez2018task}. The complex network map of the human brain, which is defined as the human connectome by \citet{Sporns_FC}, is a collection of connections in the brain. It consists of scattered voxels or regions of interest (ROIs) and their entangled spatiotemporal relationships, including structural and functional connectivity \citep{contreras2015structural}. Functional connectivity (FC) is inferred using neural activity signals such as BOLD sequences \citep{smith2013functional}. Specific statistical measures such as correlation and mutual information are commonly used to characterize the brain network structure \citep{Jeong_EEG_MI}, where a stationary interaction structure is assumed in FC calculation. However, fluctuations of both BOLD sequences and FC have become of great interest with the advancement of technology and improved experimental designs \citep{Preti_dFC}.  \citet{Vince_Review} extended the connectome concept with a dynamic perspective, namely ``dynamic connectome",  to describe a spectrum of approaches aiming to identify time-varying properties of dynamic functional connectivity (dFC). 
\subsection{Related Works}
The applications of a number of dynamic graph estimation methods developed under different assumptions have successfully identified biologically meaningful dynamic patterns for different brain states \citep{Betzel_Resting, Gonzalez_Task, Wang_Reading}. A sliding window approach is commonly adopted to link functional brain dynamics and cognitive abilities where the functional connectivity at a specific stage is constructed with observations in the corresponding time window \citep{elton2015task, fong2019dynamic}. Using the sliding window approach, \citet{sakouglu2010method} revealed different dynamic connectivities in schizophrenia in terms of task-modulation. Although the sliding window approach is straightforward and easy to interpret, the choice of the window length is still subject to debate \citep{hutchison2013dynamic, Preti_dFC}. \citet{Lindquist_bivariate} proposed multivariate volatility models to refine the sliding window methods with exponential weights and used dynamic conditional correlation to measure dFC \citep{Preti_dFC}. \citet{Chang_time_frequency} applied time-frequency analysis to circumvent the window limitation, and extended the dynamic connectivity to multiple frequencies. 

Furthermore, with the advancement of change point detection methods, BOLD sequences can be further divided into segments for a better understanding of network structure within each segment \citep{Xu_Change_DCR}. A general comparison between time segments can be achieved by a summary statistic, for example, a likelihood ratio statistic \citep{Barnett_Change_Point}, where a multivariate Gaussian distribution is assumed to facilitate statistical inference. However, it is hard to impose specific hypotheses on the heterogeneity of the graphs along a time course because changes can be continuous, which invalidates the time segments proposed by the likelihood ratio test. In \citep{park2018dynamic}, a two-sample test based on covariance matrix estimators on different time windows was proposed to detect the change time points of brain connectivity. \citet{avanesov2018change} also addressed the multi-scale properties in their testing framework, which directly tests the difference of covariance structures. Though the testing framework is similar, our method is different from the existing literature in three aspects: 1. We assume continuity instead of structural invariance for the null hypothesis, with a focus on discontinuity with sudden changes. 2. We consider the temporal dependence of the covariance structures by applying a time kernel. 3. We accordingly apply the Gaussian multiplier bootstrap to approximate the distribution of the test statistic to detect potential changes over the whole time course, and we prove that the testing procedure appropriately controls the type I error rate.

\subsection{Time-varying Gaussian Graphical Models}
Different testing statistics may be applied under similar frameworks. Given a time series $\vecX$ in a temporal scale observed on $[0, T]$, which is commonly assumed as Gaussian at a specific time point, a test statistic can be constructed for the null hypothesis where there is no change in the distribution or the covariance structure over the time course. The null hypothesis is then rejected when the test statistic derived from the observed time series reaches certain statistical significance. We  adapt the inter-subject time-varying Gaussian graphical models proposed by \citet{tan2019estimating} to consider individuals instead of two representative time series for the estimation.  The following is the setting of the inferential framework.
 Let $\vecX \in \mathbb{R}^p$ be a $p$-dimensional random variable and define a time variable $\text{T}\in[0, 1]$. In the time-varying Gaussian graphical model, we assume
$\vecX | \text{T} = t \sim N_p(\mathbf{0}, \vs(t)),$
where the covariance matrix $\vs(t)$ is defined at $\text{T}=t$ and the corresponding precision or inverse covariance is  
$\vt(t) = (\vs(t))^{-1}$
which governs the conditional dependencies for the graph. The inverse covariance matrix $\vt(t)$ is regarded as a representation of the time-varying graph $G(t)$. Thus, graphs across time can be retrieved by estimating $\vt(t)$. A kernel smoothed covariance estimator \citep{Yin_Nonpara_Cov, TVGLASSO} is used to obtain $\widehat{\vs}(t)$:
\begin{align}\label{def: GGM cov}
\widehat{\vs}(t) = \frac{\sum_{i\in[n]}K_h(t_i - t) \textbf{X}_i\textbf{X}_i ^T}{\sum_{i\in[n]}K_h(t_i - t)}
\end{align}
where $K_h(t_i - t) = K((t_i-t)/h)/h$ with $h>0$ as the bandwidth parameter. For simplicity, we use the Epanechenikov kernel described in \citet{tan2019estimating}, i.e.,
$$K(u) = 0.75\cdot(1-u^2) \mathbbm{1}_{|u|\leq 1}, $$ 
which is commonly used in non-parametric estimation \citep{epanechnikov1969non}.
The precision matrix can then be estimated using the CLIME estimator \citep{CLIME}, i.e., 
$
\widehat{\vt}_j(t) = \argminB_{\bm{\theta}\in R^p} ||\bm{\theta}||_1$, subject to $||\widehat{\vs}(t)\bm{\theta} -\textbf{e}_j||_{\infty}\leq \lambda$, 
where $\textbf{e}_j$ is the $j$-th canonical basis in $R^p$ for each $j\in\{1,...,p\}$ and $\lambda > 0$ is a tuning parameter to control the sparsity of $\widehat{\vt}_j$.
Instead of taking precision as the parameter of interest, \citet{Neykov_debiased} derived a de-biased estimator for the CLIME estimator
\begin{align}\label{def: de-biased}
\widehat{\vt}_{jk}^d(t) = \widehat{\vt}_{jk}(t) - \frac{\big (\widehat{\vt}_j(t)\big)^T\big[\widehat{\vs}(t)\widehat{\vt}_k(t) -\textbf{e}_k\big] }{\big (\widehat{\vt}_j(t)\big )^T\widehat{\vs}_j(t)},
\end{align}
which is proven to be asymptotically normal \citep{Neykov_debiased}. The construction of the de-biased precision estimator is able to circumvent the asymptotic non-normality of traditional test statistics such as score test \citep{Neykov_GPI}. 

Based on (\ref{def: de-biased}), \citet{tan2019estimating} proposed a combinatoric inferential framework to test monotone graph property for time-varying graphs. They introduced a test statistic and a bootstrap method for hypothesis testing with an appropriate type I error control.  In their work, \citet{tan2019estimating} defined the test statistic for the graph as:
\begin{align}\label{eq:tantest}
U_E &= \sup_{t\in(0, 1)}\max_{(j,k)\in E(t)} \sqrt{nh}\big |  \widehat{\vt}_{jk}^{d}(t) - \vt_{jk}(t) \big |.
\end{align}Based on the maximal statistic as defined in (\ref{eq:tantest}), we replace precision components with the difference of limits to build the test statistic for hypothesis testing. The bootstrap statistic is adapted accordingly, and the testing procedure remains the same as that in \citet{tan2019estimating} except for the skip-down operation. Finally, we introduce a testing framework called dynamic Graphical Change Inference (dGCI), which aims to test whether there is any sudden change in the covariance structure in the observed time series. 

\subsection{Our Contributions}
The major contributions of our current work are:
1. We extend the null hypothesis to allow continuous changing patterns.
2. We propose a generalized testing framework for change point detection of dynamic covariance structure.
3. Besides change point detection, we introduce a powerful tool to identify connectivity changes with adequate sample sizes.

\section{METHODS}\label{sec: test}
In this section, we introduce the time-varying Gaussian graphical models and propose testing procedures for  change point detection. We adapt model settings in \citet{tan2019estimating} and design a novel test statistic for detecting change points. We also propose a normalized estimate of the difference between precision matrices to enhance the performance of our method. 

\subsection{Test Statistic for Graph Change}

Here, we formulate the testing problem and propose a corresponding bootstrap method for statistical inference adapted from \citep{tan2019estimating}. Furthermore, we apply normalization to de-biased estimates and use normalized estimates to derive the test statistic.

\subsubsection{Inference of Graph Change}
Given the precision matrix $\vt(t)$, we note that its left and right values at $t_0$ are defined as $ \vt^{\Minus}(t_0)\!\coloneqq\!\lim_{t \rightarrow t_0-} \vt(t)$ and  $ \vt^{\Plus}(t_0)\!\coloneqq\! \lim_{t \rightarrow t_0+} \vt(t)$. We define $\vs^\Plus(t)$, $\vs^\Minus(t)$, $G^\Plus(t)$ and $G^\Minus(t)$ based on $\vs(t)$ and $G(t)$ similarly. We say edge $(j,k)$ is changed at time point $t_0$ if the left and right values of $\vt_{jk}(t)$ at $t_0$ are different, i.e., $\vt_{jk}^\Plus(t) \neq \vt_{jk}^\Minus(t)$.  Here, we do not assume temporal segments of constant precisions, which is different from the common null hypothesis with a static covariance structure \citep{avanesov2018change, Xu_Change_DCR}.\! For $G(t)\!=\!(V, E(t))$, the hypothesis testing of graph changes can be formulated as:
\begin{align}\label{def: test}
\begin{split}
&H_0\!:\!\forall t\in(0, 1),  G^\Plus(t)\!=\!G^\Minus(t),\\
&H_1\!:\!\exists t_0\!\in\!(0,1),\! (j,k)\!\in\!E(t)\! \text{ s.t. }\! G_{jk}^\Plus(t_0)\!\neq\!G_{jk}^\Minus(t_0), 
\end{split}
\end{align}
Unlike a general graph property $\mathcal{P}$ defined by \citet{Neykov_GPI} and \citet{tan2019estimating}, this specific null hypothesis does not require a skip-down procedure because the rejection of a single edge will reject the null at the same time. It simplifies the inference although we can still perform the skip-down approach for specific changes of properties. In order to observe changes of $\vt(t)$ at $\text{T}=t$, we need to consider both sides of $t$, i.e., $\vt^\Plus(t)$ and $\vt^\Minus(t)$, which represent the right and left limits, respectively. If there is no change in the current graph, it is expected that $|\vt^\Plus(t) - \vt^\Minus(t)| = 0$.

To perform the statistical test at $\text{T}=t$, first, we generalize (\ref{def: GGM cov}) into right and left approximated estimates. We define the right-sided covariance estimator as  (\ref{def: GGM cov}) with $K_h$ = $K_{h^+}$, 
where $ K_{h^+}(u) = 2K_h(u)\mathbbm 1\{u > 0\}$ to ensure $\int K_{h^+}(u)du=1$. We also define the left-sided covariance estimator $\widehat{\vs}^-(t)$ similarly with $ K_{h^-}(u) = 2K_h(u)\mathbbm 1\{u < 0\}$.

Following the same procedure above, $\widehat{\vt}^\Plus(t)$ and $\widehat{\vt}^\Minus(t)$ are estimated by GLASSO \citep{GLASSO} or CLIME \citep{CLIME} using $\widehat{\vs}^\Plus(t)$ and $\widehat{\vs}^\Minus(t)$, respectively. Similar to (\ref{def: de-biased}), the de-biased estimators of approximated estimates for the two sides are
\begin{align*}
\widehat{\vt}_{jk}^{d\pm}(t) = \widehat{\vt}_{jk}^\pm(t) - \frac{\big (\widehat{\vt}_{j}^\pm(t)\big)^T\big[\widehat{\vs}^\pm(t)\widehat{\vt}^\pm_k(t) -\textbf{e}_k\big] }{\big (\widehat{\vt}^\pm_j(t)\big )^T\widehat{\vs}^\pm_j(t)},
\end{align*}
where the notation $\pm$ means we use the left or right estimates respectively in the definition of $\widehat{\vt}_{jk}^{d+}(t)$ or $\widehat{\vt}_{jk}^{d-}(t)$.
Here, we extend (\ref{eq:tantest}) with left- and right-sided estimates for characterizing graph changes
\begin{align}\label{eq:ourtest}
\begin{split}
U_{jk}^{\pm}(t) &= \widehat{\vt}_{jk}^{d\pm}(t) - \vt_{jk}^\pm(t),\\
U_E &= \sup_{t\in(0, 1)}\max_{(j,k)\in E(t)} \sqrt{nh}\big | U_{jk}^{\Plus}(t) - U_{jk}^{\Minus}(t)\big |,
\end{split}
\end{align}
where $\vt_{jk}^\Minus(t)$ and $\vt_{jk}^\Plus(t)$ represent the left- and right-sided values of $\vt_{jk}(t)$, respectively.
We expand the Gaussian multiplier bootstrapper as
\begin{align}\label{def: left n right components}
M_{ijk}^{B\pm}\!(t)\!=\!K_{h^\pm}(t_i\!-\!t)\!\cdot\!\big (\!\widehat{\vt}_j^\pm(t)\!\big )^T\!\big [\mathbf{X}_i\mathbf{X}_i^T\!\widehat{\vt}^\pm_k(t)\!-\!\mathbf{e}_k\big]
\end{align}
where $t\in(0,1)$, $i\in[n]$ and $(j, k)\in E(t)$.
Similar to (\ref{eq:tantest}) and (\ref{eq:ourtest}), we can replace the estimates of the bootstrap statistics defined by \citet{tan2019estimating}, which are not side-specific, with the differences of the side-specific ones as shown in (\ref{eq:bootstat}). With two components defined in (\ref{def: left n right components}), the bootstrap statistic is defined as
\begin{align}
U_{jk}^{B\pm}(t) &=  \frac{\sum_{i\in[n]}M_{ijk}^{B\pm}(t)\xi_i}{\sum_{i\in[n]}K_{h^\pm}(t_i - t)}\label{eq:bootstat},\\
U_E^B &= \sup_{t\in\!(0,1)}\!\max_{(j,k)\in\!E(t)}\!\sqrt{nh}\!\cdot\!\Big | U_{jk}^{B\Plus}(t)  - U_{jk}^{B\Minus}(t) \Big |\label{eq:bootstatE},
\end{align}
where $\xi_i, ..., \xi_n\overset{i.i.d.}\sim N(0, 1)$. $c(1-\alpha, \mathbf{E})$, the conditional $(1-\alpha)$-quantile of $U_E^B$ given $\{(t_i, \mathbf{X}_i)\}_{i\in[n]}$ is defined as: 
\begin{align}\label{def: quantile}
\begin{split}
 \inf\{q\in \mathbb R|P(U_E^B\leq q|\{(t_i, \mathbf{X}_i)\}_{i\in[n]})\geq 1-\alpha\}.
\end{split}
\end{align}
In our implementation, we uniformly sample $t\in(0, 1)$ and approximate $U_E^B$ by taking the supreme over sampled $\mathbf{t}_B$ instead of $(0, 1)$. $U_E^B$ is sampled with $\bm{\xi}$ as defined in (\ref{eq:bootstat}) and (\ref{eq:bootstatE}). Similar to \citet{tan2019estimating}, the rejected edge set at $t$ is defined as
\begin{align*}
\mathcal{R}(t)\! =\! \big\{(j,k)\! \in\! E(t)\big|\sqrt{nh}\cdot\big|\Delta\widehat{\vt}_{jk}(t)\big| > c(1-\alpha, \mathbf{E})\big\},
\end{align*}
where $\Delta\widehat{\vt}_{jk}(t) = \big({\widehat{\vt}_{jk}^{d\Plus}(t) - \widehat{\vt}_{jk}^{d\Minus}(t)}\big)$. The null hypothesis will be rejected if there exists a $t_0 \in (0, 1)$ such that $\mathcal{R}(t_0)\neq \varnothing$. 

\subsubsection{Approximated Variance of Test Statistic}
With arbitrary changes of the graph, the variance of the de-biased estimator can vary. To avoid the dominance of unwanted variance, we normalize the estimates with  approximated variance. We approximate the variance of the de-biased estimator $\tilde{\sigma}^2\big (\widehat{\Theta}_{jk}^{d\pm}(t)\big )$ by taking the kernel weighted average of squares: 
\begin{align*}
 \frac{\sum_{i = 1}^NK_{h^\pm}(t_i - t)\Big( \big (\widehat{\vt}_j^\pm(t)\big)^T\big [\mathbf{X}_i\mathbf{X}_i^T\widehat{\vt}^\pm_k(t) - \mathbf{e}_k\big]\Big)^2}{\sum_{i = 1}^NK_{h^\pm}(t_i - t)}.
\end{align*}
The normalization term $\tilde{\sigma}_{jk}(t)$ for $(j,k)\in \text{V}$ is defined as $\tilde{\sigma}_{jk}(t)=\sqrt{\tilde{\sigma}^2\big (\widehat{\Theta}_{jk}^{d\Plus}(t)\big )  + \tilde{\sigma}^2\big (\widehat{\Theta}_{jk}^{d\Minus}(t)\big )}$. With $\delta\widehat{\vt}_{jk}(t) = \big({\widehat{\vt}_{jk}^{d\Plus}(t) - \widehat{\vt}_{jk}^{d\Minus}(t)}\big)/{\tilde{\sigma}_{jk}(t)}$, we normalize the test statistic
\begin{equation}\label{def: normalized test}
U_E = \sup_{t\in(0,1)}\max_{(j,k)\in E(t)}\sqrt{nh}\cdot\Big | \delta\widehat{\vt}_{jk}(t)\Big | 
\end{equation}
and the bootstrapper as
\begin{equation}\label{def: normalized bootstrapper}
U_E^B\!=\! \sup_{t\in\!(0,1)}\!\max_{(j,k)\in\! E(t)}\!\sqrt{nh}\cdot\tilde{\sigma}^{-1}_{jk}\!(t)\bigg | U_{jk}^{B\Plus}\!(t)-U_{jk}^{B\Minus}\!(t)
\bigg | 
\end{equation}
The definition of the quantile $c(1-\alpha, \mathbf{E})$ defined in (\ref{def: quantile}) remains the same. 
\subsection{Hypothesis Testing}
We summarize the bootstrap procedure and the hypothesis test in Algorithm \ref{algorithm: bootstrap} and Algorithm \ref{algorithm: testing}.
\begin{algorithm}\label{algorithm: bootstrap}
\SetAlgoLined
\textbf{Input}: Observed time sequence with time: $\mathbf{X}\in \mathbb{R}^{n\times p}$ and $\vect \in (0, 1)^{n}$; $\widehat{\Theta}^{\Plus}(t)$, $\widehat{\Theta}^{\Minus}(t)$ for $t\in \vect$; Bootstrap number $B$; $\alpha$ in the quantile defined as (\ref{def: quantile}).  \\
\While{$B >0$}{
Draw $\bm{\xi}$ from $N(\mathbf{0}, I_n)$.\\
 \For{${t \in (0, 1)}$}{
 \begin{enumerate}
 \item Calculate the normalization term $\tilde{\sigma}(t)$.
 \item Apply $\tilde{\sigma}(t)$ to the bootstrap statistic \\defined in (\ref{def: normalized bootstrapper}) with respect to $\bm{\xi}$ and \\take the maximum over $\mathbf{E}$.
 \end{enumerate}
 }
 Calculate $U_E^B$.\\
$B\gets B-1$
 }
 Calculate $c(1-\alpha, \mathbf{E})$ based on ${\mathbf{U}}_E^B$.\\
 \textbf{Output}:  $c(1-\alpha, \mathbf{E})$; $\widehat{\Theta}^{\Plus}(t)$, $\widehat{\Theta}^{\Minus}(t)$, $\tilde{\sigma}(t)$ for $t\in(0, 1)$
  \caption{Gaussian Multiplier Bootstrap for $U_E$}
\end{algorithm}

\begin{algorithm}\label{algorithm: testing}
\SetAlgoLined
\textbf{Input}: $c(1-\alpha, \mathbf{E})$;$\widehat{\Theta}^{\Plus}(t)$, $\widehat{\Theta}^{\Minus}(t)$, $\tilde{\sigma}(t)$ for $t\in(0,1)$;\\
\For{$t\in(0, 1)$}{
\begin{enumerate}
\item Compute $\delta\widehat{\Theta}(t)$
\item $\mathcal{R}(t) = \Big\{ (j, k)\in E(t)\big|\sqrt{nh}\cdot|\delta\widehat{\vt}_{jk}(t)| > c(1-\alpha, \mathbf{E})\Big\}$.
\item Reject $H_0$ if $\mathcal{R}(t) \neq \varnothing$.
\end{enumerate}
}
 \textbf{Output}:  $\mathbbm{1}\{H_0\ \text{rejected}\}$; $\mathcal{R}(t)$ for $t \in (0, 1)$.
  \caption{Hypothesis test for dynamic graph change}
\end{algorithm}

\section{THEORETICAL RESULTS}
In this section, we establish the uniform rates of convergence for the kernel smoothed covariance matrix estimator $\widehat{\vs}^\pm(t)$ and the inverse covariance estimator $\widehat{\vt}(t)$ in CLIME. In addition, we show that our testing procedure in Algorithm \ref{algorithm: testing} is a uniformly valid test. We study the asymptotic regime in which both $n$ and $p$ are allowed to increase. Our theoretical results are based on the results in \citet{tan2019estimating}, and we address the differences here and in the supplementary notes. To detect ``sudden changes'', we don't assume that the covariance matrix $\Sigma\left(\cdot\right)$ is continuous. Consequently, we obtain different rates of convergence from those in \citet{tan2019estimating}. Throughout our proof, we consider the parameter space
\begin{equation*}
\begin{split}
\mathcal{U}_{s,m,M} &=\{\Theta\in\mathbb{R}_{p\times p}| 
 \Theta \succ 0, \lambda_1(\Theta)>m, \\
& \|\Theta\|_2\leq\rho, \max_{j\in[p]}\|\Theta_j\|_0\leq s, \max_{j\in[p]}\|\Theta_j\|_1\leq M \},
\end{split}
\end{equation*}
where $0<\lambda_1(\Theta)\leq\lambda_2(\Theta)\leq\ldots\leq\lambda_p(\Theta)$ are the eigenvalues of $\Theta$. A similar class of matrices were considered in the literature on inverse covariance matrix estimation \citep{CLIME}. We allow $s$ to increase with $n$ and $p$. Next, we impose some conditions on the regularity and smoothness of the marginal density of $t$, $f_T(\cdot)$ and covariance matrix $\Sigma(\cdot)$.

\begin{assumption}\label{ass:1}
	Assume that there exists a constant $\underline{f}_T$ such that $\inf_{t\in[0,1]} f_T(t) \geq \underline{f}_T > 0$. Furthermore, assume that $f_T$ is twice continuously differentiable and that there exists a constant $\overline{f}_T < \infty$ such that $\max\{\|f_T\|_\infty, \|\dot{f}_T\|_\infty, \|\ddot{f}_T\|_\infty\} \leq \overline{f}_T$.
\end{assumption}

	Now, we introduce Assumption \ref{ass:2}. Different from Assumption 2 in \citet{tan2019estimating}, to detect graph changes, we assume $\Sigma_{jk}(t)$ is right continuous in $[0,1]$.
\begin{assumption} \label{ass:2}
	Assume $\Sigma_{jk}(t)$ is right continuous and $\Sigma_{jk}^\pm(t)$, $\dot{\Sigma}_{jk}^\pm(t)$ and $\ddot{\Sigma}_{jk}^\pm(t)$ exist for $\forall t \in (0, 1)$, $j,k \in [p]$. $\Sigma^+_{jk}(0) = \Sigma_{jk}(0)$,  $\Sigma^-_{jk}(1) = \Sigma_{jk}(1)$ and there is a finite number of discontinuities of $\Sigma(t)$ for $t \in [0,1]$. There exists a constant $M_\sigma$ such that 
\begin{equation*}
\begin{split}
	\sup_{t\in[0, 1]}\max_{j, k \in [p]}\max\{
	&\Sigma_{jk}^+(t), \dot{\Sigma}_{jk}^+(t), \ddot{\Sigma}_{jk}^+(t), \\
	&\Sigma_{jk}^-(t), \dot{\Sigma}_{jk}^-(t), \ddot{\Sigma}_{jk}^-(t)\} \leq M_\sigma.
\end{split}
\end{equation*}
\end{assumption}
	
Details of corollaries and proofs can be found in supplementary notes. Here we only state the theorems. The following Theorem \ref{theo:1} establishes the uniform rates of convergence for the kernel smoothed covariance matrix estimator under the maximum norm. The proof of Theorem \ref{theo:1} is a generalization of Theorem 1 in \citet{tan2019estimating}. Right and left Epanechnikov kernels are used in our paper to detect change points, which makes our rates of convergence different from those in \citet{tan2019estimating}. Corollary 1 in the supplementary notes gives the uniform rates of convergence of $\widehat{\vt}^\pm(t)$ defined in CLIME.
\begin{theorem}\label{theo:1}
	Assume that $h\!=\!o(1)$ and that $\log^2\!np\cdot\log (np/\sqrt{h})/nh\! =\!o(1)$. For any $0\! <\! a\! <\! b\! < \!1$, under Assumptions 1-2, there exists a positive constant $C$ such that
	$$
	\sup_{t\in [a,b]}\|\widehat{\vs}^\pm(t)-\vs^\pm(t)\|_{max} \leq C\cdot\left(h + \sqrt{\frac{\log(np/\sqrt{h})}{nh}}\right)
	$$
	for a sufficiently large $n$.
	\end{theorem}

Recall from (\ref{def: quantile}) the definition of $c(1-\alpha, \textbf{E})$. The following Theorem \ref{theo:2} shows that the Gaussian multiplier bootstrap after normalization can be approximately dominated by the quantile of the test statistic $U_E$. So it is valid by effectively controlling type I error. Similar to Theorem 2 in \citet{tan2019estimating}, to prove that $U_E^B$ is a good approximation of $U_E$, we define a series of intermediate processes, and we show both $U_E^B$ and $U_E$ can be approximated well by these processes with applications of  Theorems A.1 and A.2 in \citet{Bootstrap2}.

\begin{theorem}\label{theo:2}
	Assume that $\sqrt{nh^3} =o(1)$. In addition, assume that $\poly(s)\cdot\sqrt{\log^4(np/\sqrt{h})/nh^2} + \poly(s)\cdot\log^8(p/h)\cdot\log^2(ns)/(nh)=o(1)$, where $\poly(s)$ is a polynomial of $s$. Under the same conditions in Corollary 2 of supplementary notes, we have
	$$
	\lim_{n\rightarrow\infty}\sup_{\vt(\cdot)\in\mathcal{U}_{s,m,M}}P_{\vt(\cdot)}\left(U_E \geq c\left(1-\alpha, E\right)\right) \leq \alpha.
	$$
	\end{theorem}
\textbf{Remark.} By integrating the scaling conditions that $\sqrt{nh^3} =o(1)$ and $1/\left(nh^2\right)=o(1)$, we conclude that the rate of $h$ is $n^{-1/2}\precsim h \precsim n^{-1/3}$, where we say $a_n\precsim b_n$ when $a_n, b_n \geq 0$ and there exists a constant $C$ such that $C\cdot a_n \leq b_n$ for any $n=1,2,\ldots$. In practice, we set $h \sim n^{-0.4}$, which satisfies the scaling conditions.
\section{SIMULATION}

We evaluate the performance of our proposed method through simulations under various scenarios. Each simulation consists of 500 repeated runs with specific parameters including the number of observed time points $N_T$, the number of time segments with different precisions and the number of nodes $p$. Then, we uniformly generate $N_T$ time points $t\in(0, 1)$ along with $\vt(t)$. For the estimation of precisions, we use the CLIME estimator at each time point. In all simulations, we set $h\! =\! C_1\cdot n^{-0.4}$ and $\lambda\! =\!C_2\cdot( h\!+\!\sqrt{log(np/\sqrt{h})/(nh)})$, where $C_1$ and $C_2$ are the tuning parameters. We uniformly take 50 points in $(0, 1)$ as $\mathbf{t}_B$ for bootstrapping, and set $\alpha\! =\! 0.05$ for hypothesis testing.

\subsection{Changing Precision without Sudden Breaks}
To test if the method can control type I error rate when there is no sudden change, we simulate time series with smoothly changing precisions. We first generate $N_U=100$ sparse integer vectors $\mathbf{U}$ using a multinomial distribution with $p = \{50, 100\}$ elements with equal probability, and set the number of trials $m = 3$. The graph $\vt$ is generated as $\sum_{i\in[{N_U}]} \beta_i \cdot\mathbf{U}_i \mathbf{U}_i^T$ where $\bm{\beta}$ is the set of coefficients drawn from $\text{Unif}(0, 1)^{N_U}$. We randomly flip the signs of entries in $\vt$ to introduce negative entries. To introduce continuously and smoothly changing pattern of precisions, we generate $\{\vt_1, \vt_2\}$, and $\vt(t)$ is defined as the linear interpolation of these two precision matrices given $t\in(0, 1)$.  We set $C_1 = 1$ when $p = 50$, and $C_1 = 2$ when $p = 100$. $C_2$ is set to $0.2$ which can be selected by cross-validation for CLIME. 
\begin{table}[h]
\centering
\caption{Type I error rate}
\begin{tabular}{rrrrr}
  \hline
 &$N_T$=$700$ & $800$ & $900$ & $1000$ \\ 
  \hline
$p$=50 & 0.04 & 0.04 & 0.04 & 0.05  \\ 
100 & 0.04 & 0.05 & 0.04 & 0.04  \\ 
   \hline
\end{tabular}
\label{tab: type1}
\end{table}
As shown in Table \ref{tab: type1}, type I error is controlled under $\alpha = 0.05$. We also compare our method with covcp introduced by \citet{avanesov2018change} and Dynamic Connectivity Detection (DCD) proposed by \citet{Xu_Change_DCR}. We use the same parameter settings described in \citet{avanesov2018change, Xu_Change_DCR} for testing, in which the precision estimator with the same parameters used in our simulations is applied for covcp. Different from our approach, both these methods assume a constant $\vt$ throughout the time course. Since the covariance structure is changing through the time course, both covcp and DCD rejected the null hypothesis over 90\% of all simulations.

\subsection{Changing Precision with Sudden Breaks}
Within a diagonal matrix $G_p = \text{diag}(g_1, ..., g_p)$, $\vt_{ij}$ in $M$ entries are assigned with $a\cdot g_i g_j$ in which $0 < a < 1$. The signs are randomly set to introduce negative entries. After the initialization of $\vt$, we set $\vt_{jj} = \vt_{jj} - \min(\Lambda_{min}, 0)+0.05$ where $\Lambda_{min}$ is the minimum of eigenvalues of $\vt$. We generate one matrix with random entries and randomly flip the signs to form $(\vt_1^{(1)}, \vt_1^{(2)})$, $(\vt_2^{(1)}, \vt_2^{(2)})$ and $(\vt_3^{(1)}, \vt_3^{(2)})$. The precision is a linear function of $T$ for each segment, which is an interpolation of each matrix pair. The diagonal precision entries are randomly initialized with $1$ or $9$ with the ratio 9:1. We set $p = 50$; $M = 50$; $N_T=\{1000, 1500, 2000, 2500, 3000\}$; $a=0.2$; $C_1 = 2$; $C_2 = 0.4$.  A change point is considered successfully identified if it is localized within half of the bandwidth. We compared the power of covcp and DCD with our method for identifying both change points. As shown in Table \ref{tab: power}, dGCI has the best power of both change points with relatively small $N$ and is slightly better than DCD with larger $N$. covcp cannot effectively identify both change points even with an increasing sample size when $a=0.2$. All three methods can identify both change points with more than 85\% power when $a = \{0.3, 0.4\}$. 
\begin{table}[ht!]
\centering
\caption{Power of detecting change points with $\alpha=0.2$}
\begin{tabular}{rrrrrr}
  \hline
 & $N_T$=1000 & 1500 & 2000 & 2500 & 3000 \\ 
  \hline
dGCI & 0.40 & 0.59 & 0.76 & 0.89 & 0.95 \\ 
 DCD & 0.04 & 0.27 & 0.59 & 0.85 & 0.91 \\ 
 covcp& 0.26 & 0.17 & 0.19 & 0.24 & 0.37 \\ 
   \hline
\end{tabular}
\label{tab: power}
\end{table}

Additionally, we tested if our approach can locate changed edges besides change points. Table \ref{tab: atp} and Table \ref{tab: afp} show the increasing ability of the algorithm to distinguish sudden changes in precisions given more samples and larger effect sizes. Here, sensitivity is the ratio of successfully identified edges over changed ones, and the false discovery rate is the ratio of falsely identified edges over all the detected.

\begin{table}[ht!]
\centering
\caption{Average senstivity}
\begin{tabular}{rrrrrr}
  \hline
 & $N_T$=1000 & 1500 & 2000 & 2500 & 3000 \\ 
  \hline
a=0.2 & 0.06 & 0.08 & 0.09 & 0.12 & 0.15 \\ 
0.3 & 0.23 & 0.31 & 0.40 & 0.46 & 0.52 \\ 
0.4& 0.28 & 0.36 & 0.44 & 0.50 & 0.54 \\ 
   \hline
\end{tabular}
\label{tab: atp}
\end{table}

\begin{table}[ht!]
\centering
\caption{Average false positive rate}
\begin{tabular}{rrrrrr}
  \hline
 & $N_T$=1000 & 1500 & 2000 & 2500 & 3000 \\ 
  \hline
a=0.2 & 0.18 & 0.09 & 0.04 & 0.02 & 0.02 \\ 
 0.3 & 0.05 & 0.03 & 0.02 & 0.01 & 0.01 \\ 
 0.4& 0.06 & 0.03 & 0.03 & 0.02 & 0.02 \\ 
   \hline
\end{tabular}
\label{tab: afp}
\end{table}

\section{APPLICATION}

\subsection{Data Description}
A high-resolution fMRI dataset was collected from 15 scanned participants with an audio-video presentation of ``Forrest Gump" \citep{hanke2014high, hanke2016studyforrest}. All the subjects had listened to an audio presentation of the movie, and only one subject had never seen the movie before the experiment \citep{hanke2016studyforrest}. The original movie was cut into eight parts. Descriptions of semantic conflicts were extended from simple annotations of the cuts and scenes \citep{hanke2016lies}. With the help of scene descriptions, we can locate the scenes of BOLDs. These extended descriptions are references for FC changes.  \href{https://github.com/psychoinformatics-de/studyforrest-data-aggregate}{The fMRI data} were processed with standard methods and parcellated into 268 nodes using a whole-brain, functional atlas defined in a separate sample \citep{shen2013groupwise}. Time series within the same node were averaged, and a sequence of 3539 frames was generated per node, in which TR (repetition time) is 2s. With part of the cerebellum (9 at both hemispheres) and brainstem (1 at the left) missing in some individuals, the total number of ROI is 249. This dataset, along with detailed preprocessing methods, can be attained from ``\href{http://studyforrest.org}{http://studyforrest.org}".

\subsection{Objectives}

Individuals may share similar reactions when watching the same movie segment. We aim to capture the individual similarity in various contexts of interest. For example, certain regions of the brain may be more sensitive to specific cuts or conflicts. The changed edges identified using our inferential framework could be strengthened, weakened, and reversed in direction. The biological meaning of the identified edges is that the corresponding brain network structure may have a sudden change within a specific time or event, which could be the response of the brain to the external stimuli. Given an fMRI scan from an individual, we perform hypothesis testing, as described in section \ref{sec: test}. We are interested to see \textbf{1}: during the movie, which ROI has the most altered edges, \textbf{2}: which edge/region is mostly altered across all tests, and \textbf{3}: which edges are similar among individuals for a specific scene of interest. 

\subsection{Results}

We first scale 3539 consecutive frames into the $(0, 1)$ range and we uniformly take 30 points in $(0, 1)$ for bootstrapping. Besides these 30 points, 7 time points among 8 cuts are also included for hypothesis testing. We set $C_1 = 1.0$, $C_2= 0.4$, $\alpha = 0.05$, and the number of bootstrap at 500. To remove subject-specific intrinsic BOLDs fluctuations, we calculate the kernel smoothed BOLD sequences with the same $h$ as $\hat{\mu}(t) = \sum_{t^\prime = t_1}^{t_T}k_h(t^\prime, t)\mathbf{X}_{t^\prime} /  \sum_{t^\prime = t_1}^{t_T}k_h(t^\prime, t)$, and subtract it from the original sequences. 

\begin{figure*}
\centering
\begin{tabular}{ccc}
\includegraphics[scale=0.28]{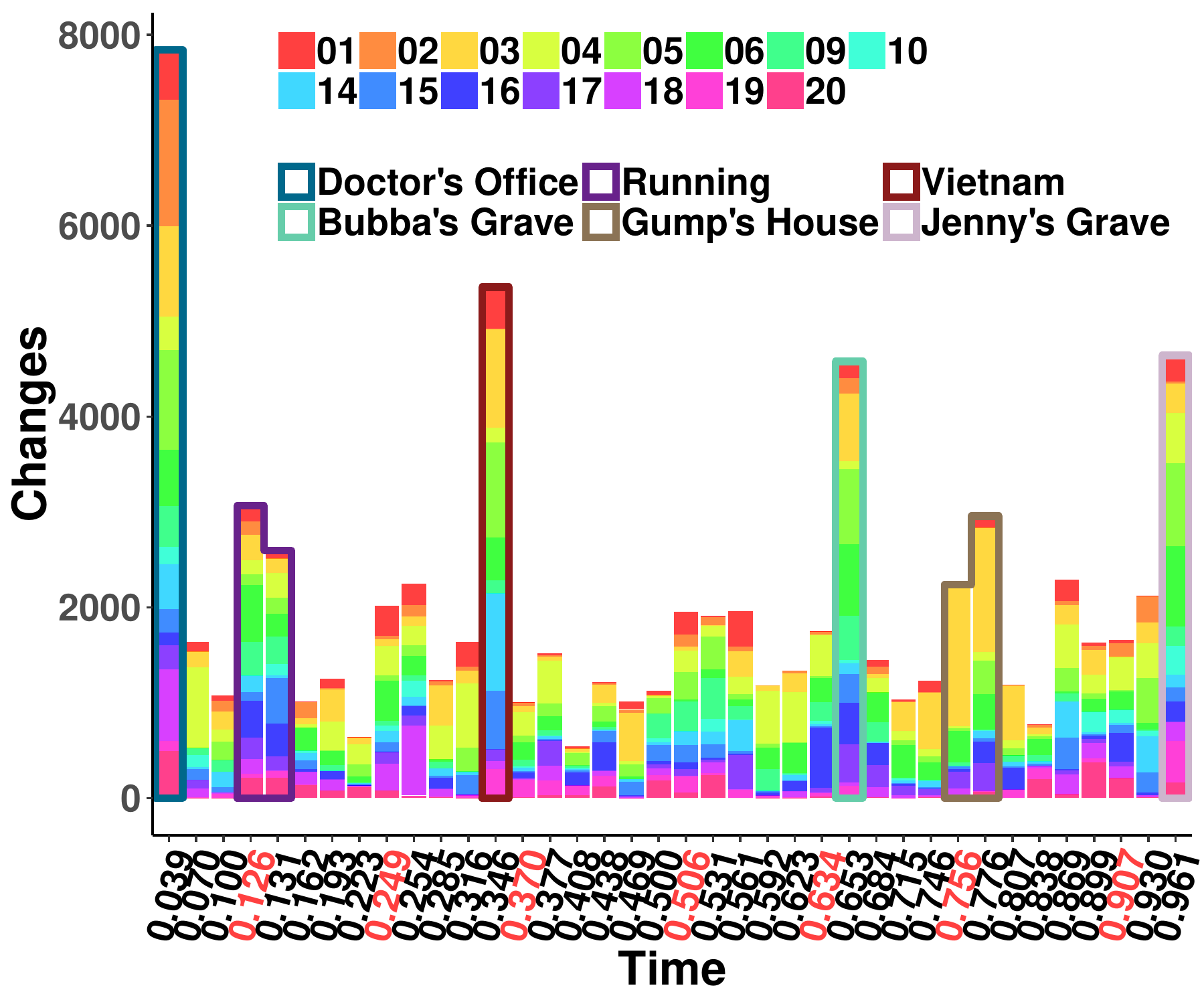}  &
\includegraphics[scale=0.28]{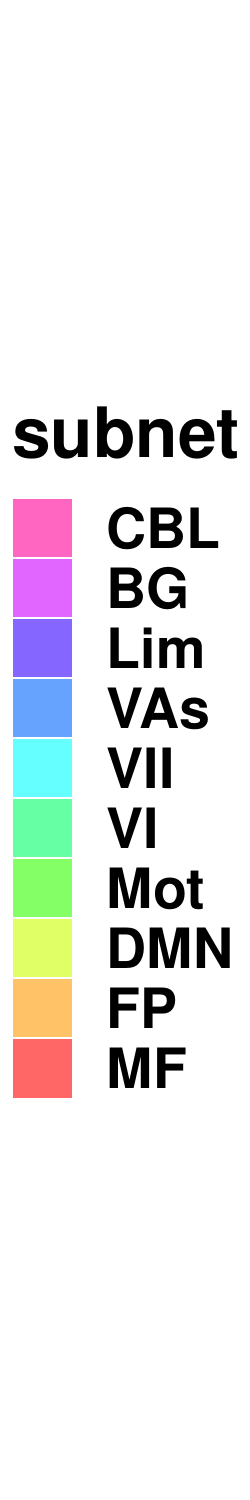}
\includegraphics[scale=0.28]{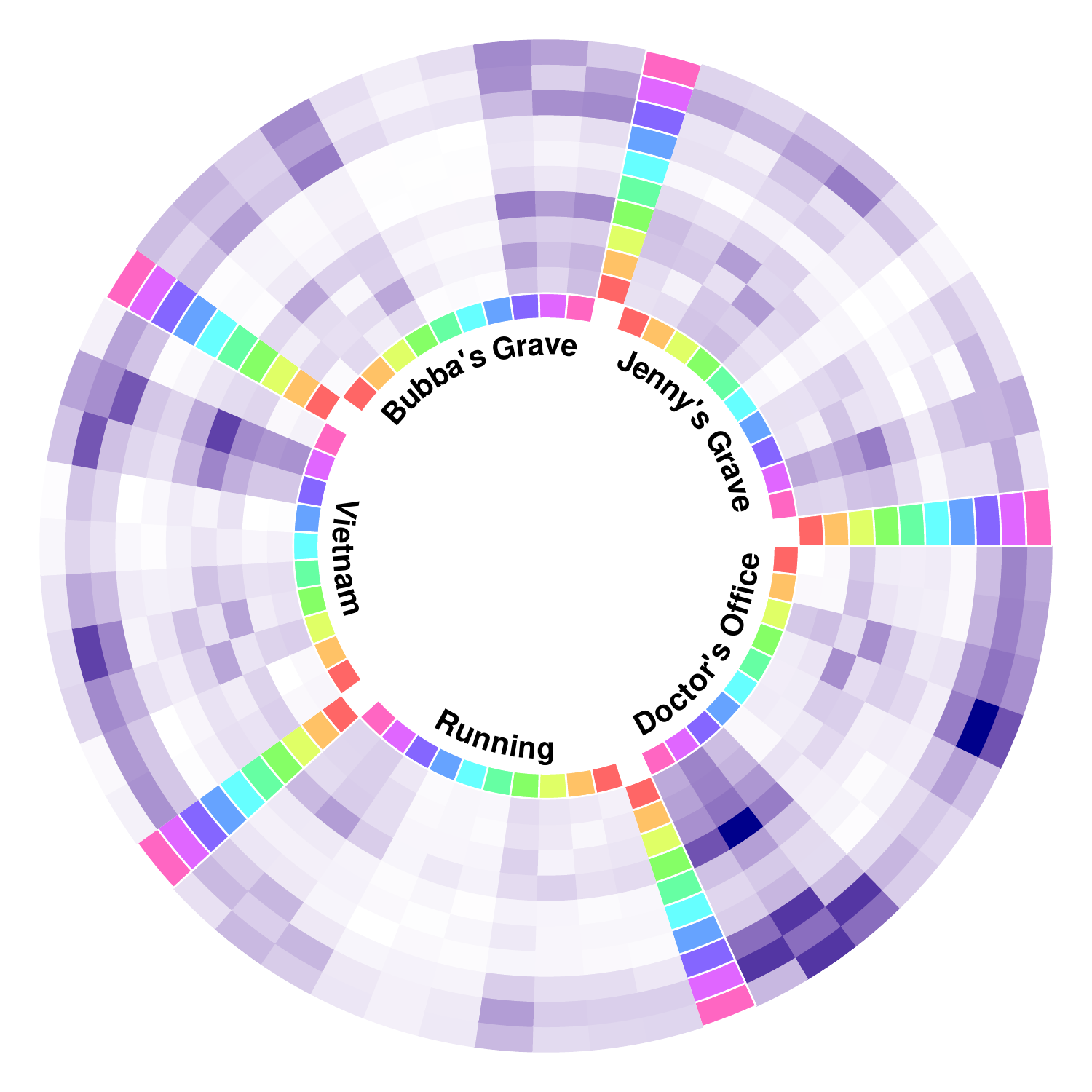}
\includegraphics[scale=0.28]{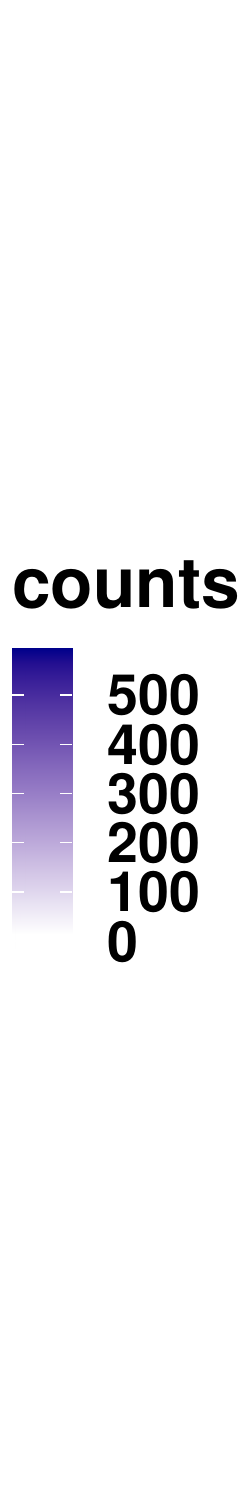} &
\includegraphics[scale=0.28]{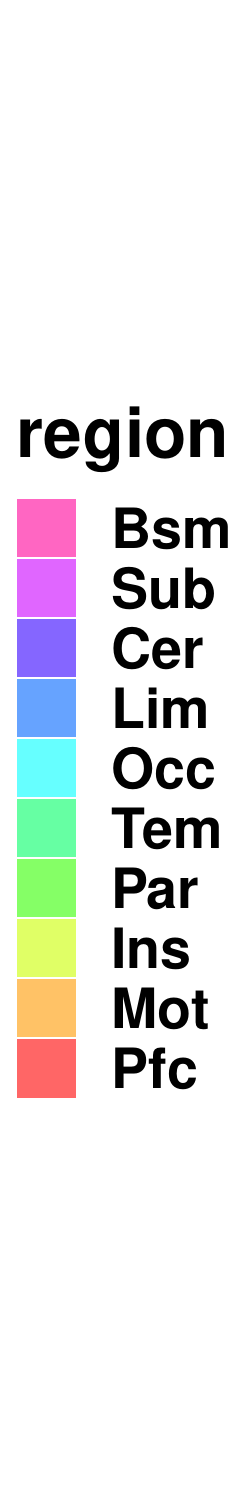}
\includegraphics[scale=0.28]{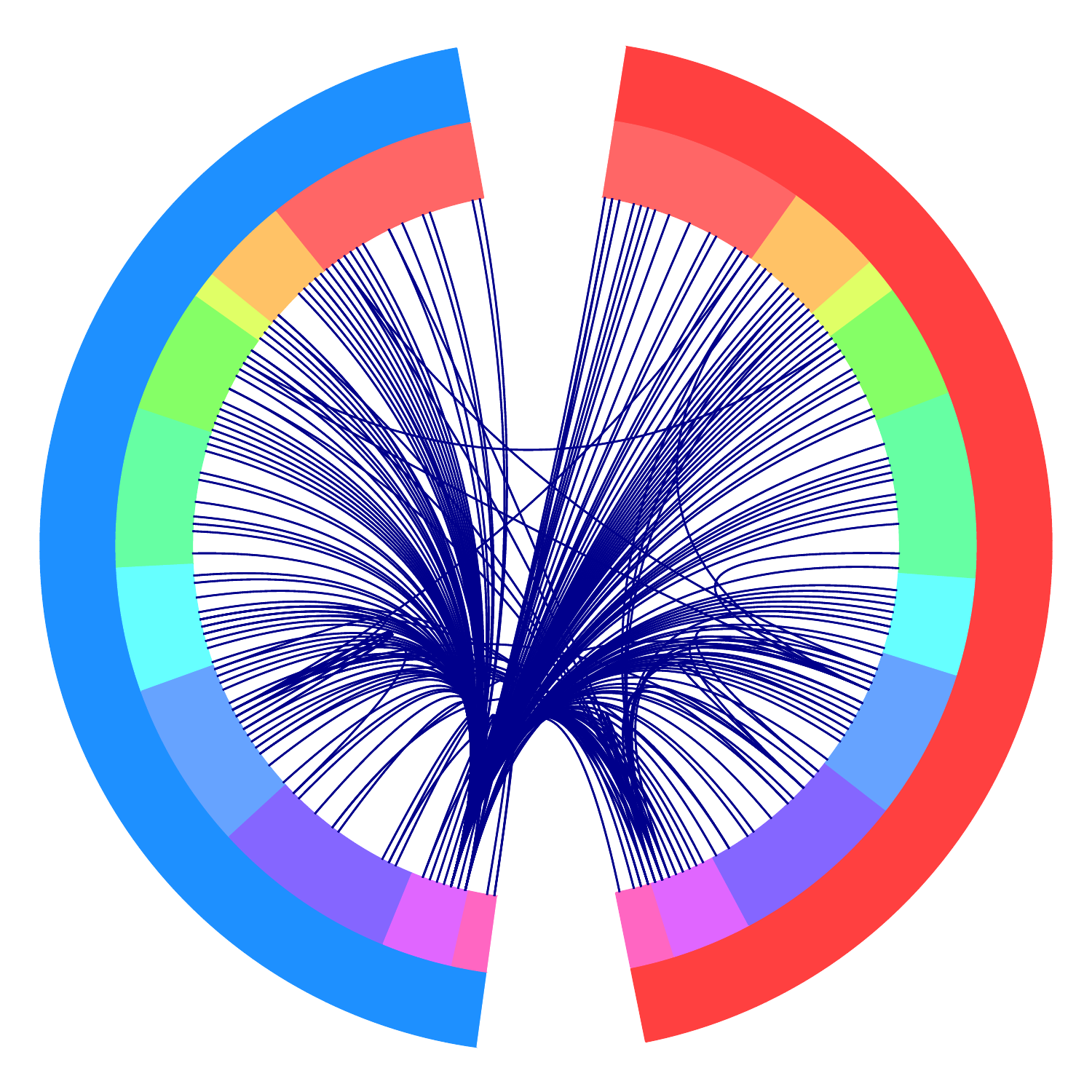} \\
(a)& (b)&(c)\\
\end{tabular}
\caption{(a) Aggregated edge number. Cuts are labeled as red texts in x-axis. Individuals are labeled in different colors in bars. (b) Counts of total identified changed edges cross subnets.  (c) Identified edges cross regions.}
\label{fig: bars}
\end{figure*}

The results from six females and nine males are aggregated as shown in Figure \ref{fig: bars}a, and the original indices for these individuals are ``01", ``02", ``03", ``04", ``05", ``06", ``09", ``10", ``14", ``15", ``16", ``17", ``18", ``19", and ``20". We highlight some scenes that stand out for the identified edge number. Notably, individual ``03", who had never watched the movie, yields more changed edges compared with the other individuals--especially during the scene at Gump's House when the main character Jenny returns. The top five scenes with the largest number of identified edge changes for both sexes are labeled as ``Doctor's Office", ``Running", ``Vietnam", ``Bubba's Grave" and ``Jenny's Grave".  ``Forrest Gump" has a considerable number of scenes in which the main character recollects previous experiences, and we only label the primary scene for a given time. These five major scenes are also identified as changed points for all the individuals regardless of sex. 

To investigate the functions of the identified changes, we group nodes with definitions of functional and anatomical annotations. The ten regions are prefrontal (PFC, nodes number: 46), motor (Mot, 21), insula (Ins, 7), parietal (Par, 27), temporal (Tem, 39), occipital (Occ, 25), limbic (Lim, 36), cerebellum (Cer, 23), subcortex (Sub, 17) and brainstem (Bsm, 8) and ten canonical brain networks include medial frontal (MF, 29), frontoparietal (FP, 28), default mode (DMN, 18), motor (Mot, 49), visual I (VI, 18), visual II (VII, 8), visual association (VAs, 17), limbic (Lim, 30), basal ganglia (BG, 29) and cerebellum (CBL, 23). \citep{smith2009correspondence, shen2013groupwise, XilinID}. We group results of five interesting scenes from Figure \ref{fig: bars}a into interactions among nodes in sub-networks. As shown in Figure \ref{fig: bars}b, limbic, basal ganglia, and cerebellum form a cluster at the important scenes. The BG network, including thalamus and striatum, also has a strong linkage with the motor network and serves as a hub network for the cortical and the subcortical regions.

To make sure the results are not dominated by outliers, we plot the connectome (Figure \ref{fig: bars}c) with edges identified within at least three individuals at the same scene. The subcortical region, which is the most critical component of the BG network, exhibits reproducible connections with different lobes from the whole brain.  It was believed that the basal ganglia structure is the primary modulator or hub for cortical information flow \citep{lanciego2012functional}. Recent studies have extended the functional scope from motor to cognitive functions \citep{helie2013exploring}. The results indicate that the basal ganglia may play a key role in processing complex stimuli like movies.

\begin{figure}[ht!]
\centering
\begin{tabular}{ccc}
Change-Emotion Correlation \\
\includegraphics[scale=0.4]{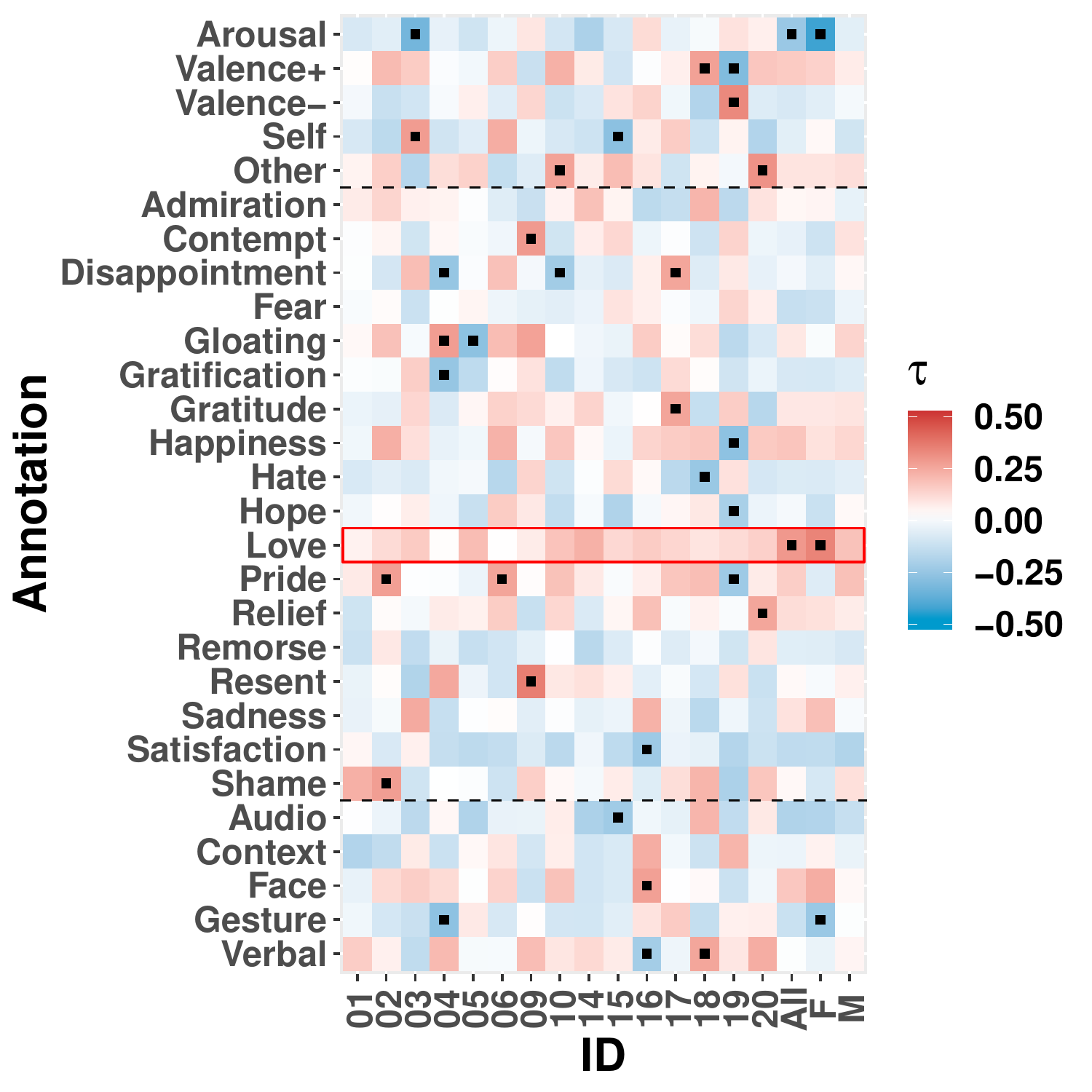} & \\
 Emotion: "Love" \\
\includegraphics[scale=0.4]{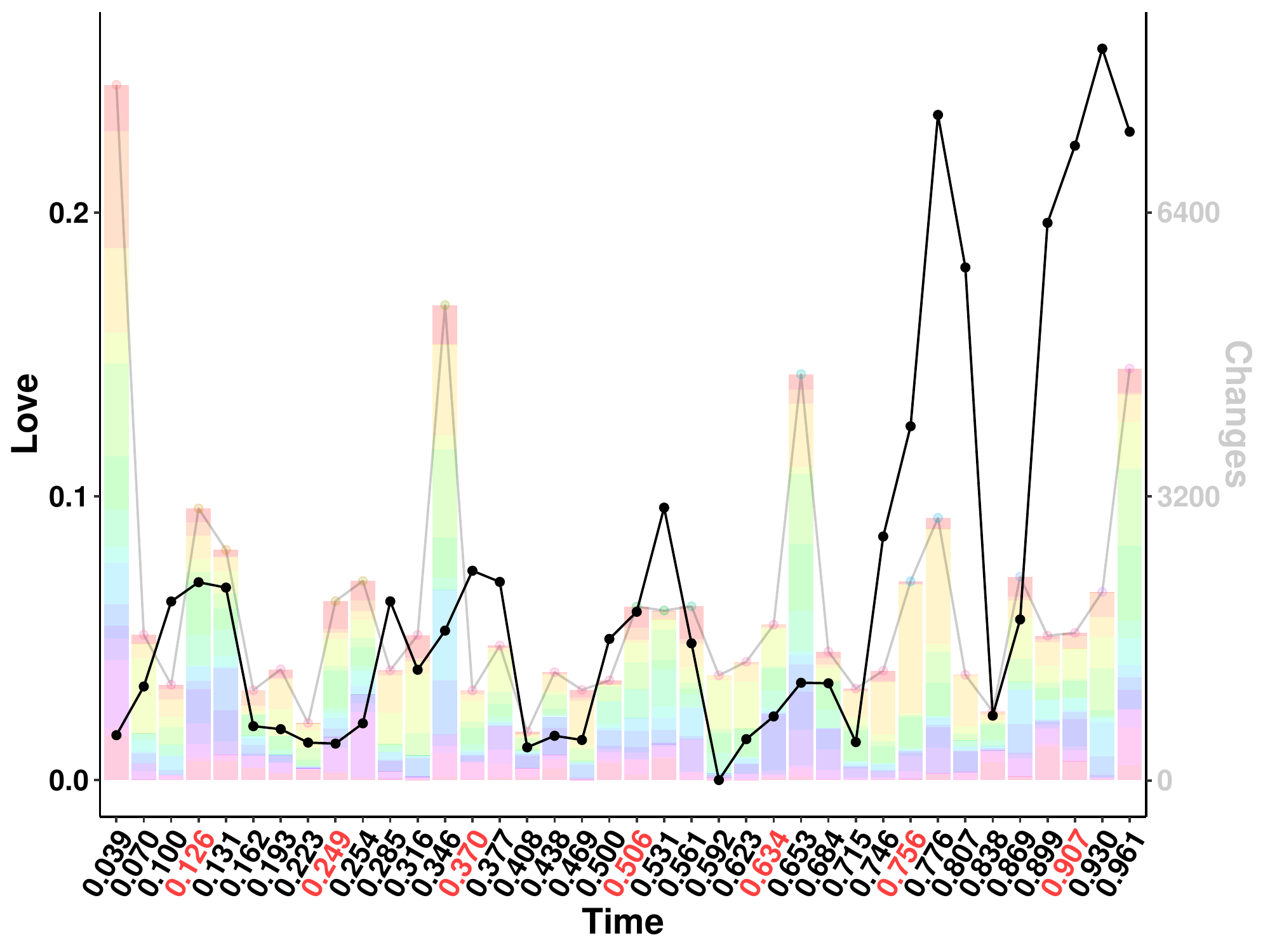}
\end{tabular}
\caption{Kendall's tau between annotations and counts of changed edges, in which grids filled with black dots have $p < 0.05$; Emotion: ``Love" overlays with changes over scenes.}
\label{fig:emotions}
\end{figure}

It is also intriguing to relate the FC response with published emotional annotations. \citet{emotions} annotated the audio-visual stimulus of ``Forrest Gump" with three major groups: dimensional emotion attributes (arousal, valence, direction), emotion categories (from admiration to shame) and emotion onset cues (audio, context, face, gesture, narrator, verbal) with detailed descriptions. To investigate if the changes of FC are correlated with any emotions, we use the same kernel in our framework to smooth the collected emotion measure within the bandwidth. We calculate Kendall's $\tau$ between emotion measures and changed edge numbers for each individual and aggregated sex group. Interestingly, we identified ``Love", which is explained as ``Affection for another person", to be positively correlated with FC changes for almost all the individuals, and it is significant in average connection change for all subjects and female individuals. The contribution of changes with the basal ganglia system may be caused by its dopaminergic role as in neurotransmitter release. Further investigation is needed to relate emotion, emotion response in FC, and emotion-related neurotransmitters.

\section{DISCUSSION}
Change point detection of dynamic functional connectome involves extensive multiple hypothesis testing. By designing the supreme-maximum test statistic based on the de-biased estimator with asymptotic normality, the hypothesis testing procedure can identify change points along with strong edge changes with appropriate false positive control. This will improve our understanding of the dynamic flow of functional connectivity given complex stimuli. Our method is able to identify psychologically meaningful changes in dynamic functional connectivity, and our results suggest the involvement of the basal ganglia in complex cognitive processes. 

We note that our method is not based on the population level modeling, which makes it difficult to aggregate the identified changes. Additionally, the method might lead to inflated type I error without enough observations when the change of connectivity is not smooth, which is not an issue of the validity of the theoretical results but an issue worth studying in applications. Despite the limitations, our algorithm can still detect consistent changes of individuals in real data.

\subsubsection*{ACKNOWLEDGEMENTS}

Here we want to express our gratitude to Dr. Han Liu from the MAGICS (Modern Artificial General Intelligible and Computer Systems) lab at the Northestern University, who provided a valuable chance to connect the authors. We are also grateful for the help of Dr. Kean-Ming Tan in the Department of Statistics at the University of Michigan, whose work inspired and motivated ours. We want to thank Dr. Xilin Shen at Yale Magnetic Resonance Research Center for providing both technical details and insights into imaging processing and analyses.

\bibliography{aistats_camera}

\end{document}